\documentclass[trackchanges,twocolumn]{aastex63}
\usepackage{CJKutf8}

\newcommand{\msolph}{~h^{-1} \rm{M}_{\odot}}
\newcommand{\Msun}{~h^{-1} \rm{M}_{\odot}}
\newcommand{\mpcph}{~h^{-1}\rm{Mpc}}

\shorttitle{Cluster-counterpart Voids in Galaxy Density Fields}
\shortauthors{Shim et al.}

\begin{document}
\begin{CJK}{UTF8}{mj}
\title{Cluster-counterpart Voids: Void Identification from Galaxy Density Field}

\correspondingauthor{Juhan Kim}
\email{kjhan@kias.re.kr}

\author[0000-0001-7352-6175]{Junsup Shim}
\affiliation{School of Physics, Korea Institute for Advanced Study, 85 Hoegiro, Dongdaemun-gu, Seoul 02455, Korea}
\affiliation{Institute of Astronomy and Astrophysics, Academia Sinica, No.1, Sec. 4, Roosevelt Rd, Taipei 10617, Taiwan}

\author{Changbom Park}
\affiliation{School of Physics, Korea Institute for Advanced Study, 85 Hoegiro, Dongdaemun-gu, Seoul 02455, Korea}

\author[0000-0002-4391-2275]{Juhan Kim}
\affiliation{Center for Advanced Computation, Korea Institute for Advanced Study, 85 Hoegiro, Dongdaemun-gu, Seoul 02455, Korea}

\author[0000-0003-4923-8485]{Sungwook E. Hong(홍성욱)}
\affiliation{Korea Astronomy and Space Science Institute, 776 Daedeokdae-ro, Yuseong-gu, Daejeon 34055, Korea}
\affiliation{Astronomy Campus, University of Science \& Technology, 776 Daedeok-daero, Yuseong-gu, Daejeon 34055, Korea}

\begin{abstract}
We identify cosmic voids from galaxy density fields under the theory of void-cluster correspondence. We extend the previous novel void-identification method developed for the matter density field to the galaxy density field for practical applications. From cosmological N-body simulations, we construct galaxy number- and mass-weighted density fields to identify cosmic voids that are counterparts of galaxy clusters of specific mass. The parameters for the cluster-counterpart void identification such as Gaussian smoothing scale, density threshold, and core volume fraction are found for galaxy density fields.
We achieve about $60$--$67\%$ of completeness and reliability for identifying the voids of corresponding cluster mass above $3\times10^{14}\msolph$ from a galaxy sample with the mean number density, $\bar{n}=4.4\times10^{-3} (h^{-1}{\rm Mpc})^{-3}$. When the mean density is increased to $\bar{n}=10^{-2} (h^{-1}{\rm Mpc})^{-3}$, the detection rate is enhanced by $\sim2$--$7\%$ depending on the `mass scale' of voids.
We find that the detectability is insensitive to the density weighting scheme applied to generate the density field.
Our result demonstrates that we can apply this method to the galaxy redshift survey data to identify cosmic voids corresponding statistically to the galaxy clusters in a given mass range.

\end{abstract}



\keywords{Large-scale structure of the universe (902), Voids (1779)}

\section{Introduction}\label{sec:Intro}
Cosmic voids are the largest volume component of the large-scale structures in the universe \citep{joeveer&einasto78,einasto+80,gott+86,lapparent+86, vogeley+94} but contain a relatively small amount of mass. Because of their underdense nature, voids have been utilized as a probe for dark energy \citep{lee&park09,lavaux&wandelt10,li11,pisani+15,sutter+15, achitouv17}, gravitation \citep{nusser+05, li+12, cai+15, achitouv19}, initial conditions \citep{kim&park98}, as well as a laboratory for studying the environmental effect on galaxy formation and evolution \citep{ceccarelli+08,park&lee09, kreckel+11,beygu+13, shim+15, ceccarelli+22}.

For void identification from simulation and observation, various void-finding methods have been developed \citep{kauffmann91,el-ad96, colberg+05, patiri06, hahn+07, platen&jones07, neyrinck08, forero09, lavaux&wandelt10,sutter+15, shim+21}.
With reasonable void-finding parameter values, a void-finder may identify voids that are aspherical and hierarchical. However, neither the asphericity nor hierarchical structure of voids
are taken into account in the spherical void formation model \citep{filmore&goldreich84, suto+84, bertschinger85}.
Thus, it is not unnatural to find some discrepancy between voids in data and the spherical void model. For example, the shell-crossing density threshold derived from the statistics of voids in data is inconsistent with the prediction based on the spherical void approximation \citep{chan+14, achitouv+15, nadathur&hotchkiss15}.
This motivates a search for a new definition and identification method for voids that mitigate the inconsistency between voids in theory and data.

Recently, a new concept of defining voids has been presented \citep{pontzen+16,shim+21,stopyra+21,desmond+22}. The key idea of void identification in these studies is that a void of a certain size can be related with a dark matter halo of a particular mass because a halo region would have become a void if the initial overdensity field were inverted. The correspondence between voids and dark matter halos in the inverted field is more reliable on the scale of massive galaxy clusters. This argument leads us to define cosmic voids as counterpart to galaxy clusters, and we will call the void of a certain size identified and related with the cluster of corresponding mass, the \emph{cluster-counterpart void} (CCV).
Other interesting features of CCVs include 1) almost universal density profiles \citep{shim+21} and 2) their mean density contrast being close to the prediction from the spherical expansion model \citep{stopyra+21,desmond+22}.
Interestingly, 3) they are still in the quasi-linear regime, and information on the initial conditions is better preserved relative to clusters \citep{kim&park98, stopyra+21}.

In order to identify CCVs from a given density field, two different approaches have been suggested. The major difference between the two approaches is whether the void identification method requires the reconstruction of the initial density fields. Without utilizing the initial density field, \citet{shim+21} established a simple CCV identification method adopting free parameters for smoothing scale, density and volume thresholds. On the other hand, reversely evolving a given density field to the initial state and then forwardly evolving its sign-inverted density fluctuation field to the present epoch to identify CCVs was suggested by \citet{stopyra+21} and implemented by \citet{desmond+22}. In this study, we extend and apply the former approach adopted in \citet{shim+21} to galaxy density fields.

This paper is organized as follows. We describe our simulations and the scheme for assigning galaxies to halos in Section~\ref{sec:simulation}. In Section~\ref{sec:review}, we briefly review the identification of CCVs from dark matter density fields and describe how we apply this approach to galaxy density fields in Section~\ref{sec:result}. We discuss our results and conclude in Section~\ref{sec:discussion and summary}.

\section{Simulation and Galaxy Mocks}\label{sec:simulation}
We use a pair of N-body simulations with inverted initial overdensity fields to study the CCVs and their corresponding clusters. The Mirror simulation used in this work is one of the Multiverse simulations introduced by \citet{park+19} and \citet{tonegawa+20}. The Reference simulation starts with the initial density fluctuations that have the same amplitude as in the Mirror simulation but have the opposite sign. The simulations adopt the GOTPM \citep{dubinski+04} code and WMAP 5--year $\Lambda$CDM cosmology \citep{dunkley+09} with the matter, baryon, dark energy density parameters set to $0.26$, $0.044$, and $0.74$, respectively. Each simulation evolves $N_p=2048^3$ dark matter particles with mass $m_p\simeq 9\times 10^9 \Msun$ in a periodic cubic box of side length $L_{\rm box}=1024\mpcph$.

We identify dark matter halos with 30 or more particles using the Friend-of-Friend (FoF) algorithm with the linking length $l_{\rm link}=0.2\overline{l_{\rm p}}$, where $\overline{l_{\rm p}}$ is the mean particle separation.
We then populate the identified halos with mock galaxies based on the most bound member particle (MBP)--galaxy correspondence model \citep{hong+16}. In this galaxy assignment scheme, all MBPs marked in halo merger trees are the proxies for galaxies \citep{delucia+04, faltenbacher&diemand06}. The position and velocity of a galaxy are taken from those of a MBP, whereas galaxy luminosity is determined by the abundance matching between the mass function of mock galaxies \citep[see][for the definition of galaxy mass]{hong+16} and the luminosity function of the SDSS main galaxies \citep{choi+07}. The survival time of a satellite halo is given by the merger time scale modeled with a modified version of the fitting formula described in \citet{jiang+08}. We set the power-law index of the host-to-satellite mass ratio to 1.5. This is to yield a good match for the projected two-point correlation functions between the mock galaxies in our simulations and SDSS main galaxies down to scales below 1$\mpcph$ \citep{Zehavi+11}.

We construct two galaxy samples with different mean number densities to test the effect of sample density on the CCV identification. The mean number densities of the sparse and dense samples are $\bar{n}_{\rm gal}=4.4\times10^{-3}$ and $1.0\times10^{-2}\ h^{3}{\rm Mpc}^{-3}$, respectively. 
We chose the sparse sample density as such assuming that the sparse galaxy sample mimics a volume-limited sample from the SDSS main galaxy sample with the largest survey volume \citep{choi+10best}.
For the dense sample, its mean density is set close to the highest achievable density from the simulations. The dense sample-like data will be available from upcoming surveys such as Dark Energy Spectroscopic Instrument \citep{DESI+16} and SPHEREx \citep{SPHEREx+14} given that their expected number densities of observed galaxies up to $z\approx0.2$ are higher than our dense sample density.

\section{Identifying CCVs with dark matter}\label{sec:review}
We briefly summarize how CCVs are defined using a paired simulations in Section~\ref{subsec:CCV} and are recovered in dark matter density fields without using the inverted simulation in Section~\ref{subsec:CCVfromDM}.

\subsection{Cluster-counterpart voids}\label{subsec:CCV}

\begin{figure*}
	\includegraphics[trim={3cm 0 22cm 0},clip,width=2\columnwidth]{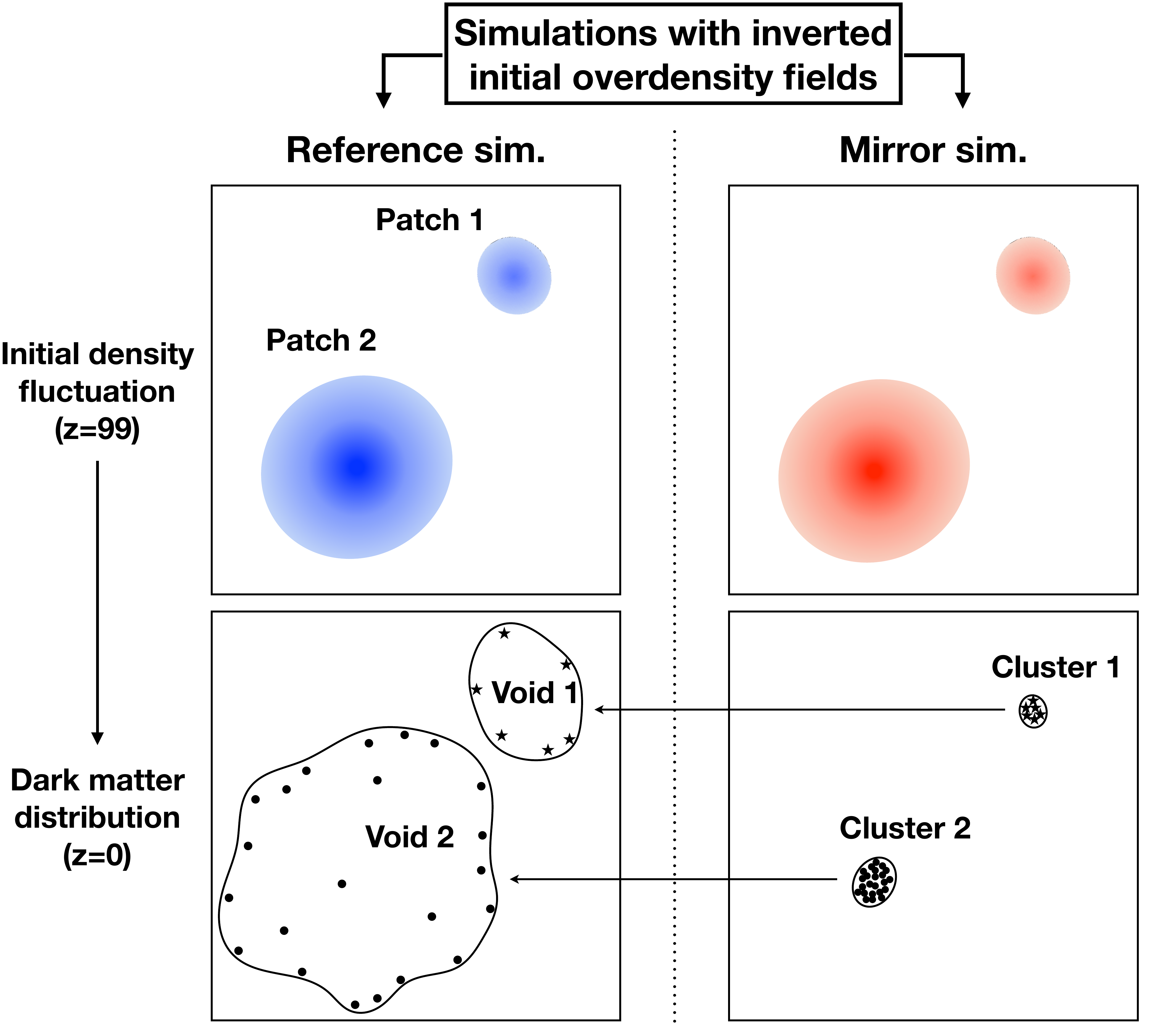}
    \caption{A schematic diagram showing how CCVs are identified under the paradigm of the void-cluster correspondence theory using the Reference and Mirror simulations. \textit{Top panels}: Initial density fluctuations of the same regions within the Reference (left) and Mirror (right) simulations. The initial density troughs (blue regions) in the Reference simulation correspond to the initial density peaks (red regions) in the Mirror simulation because the initial overdensity fields of the two simulations are designed to be the sign-inverted version of the other. \textit{Bottom panels}: Distributions of dark matter particles forming voids (left) and their counterpart clusters (right) through gravitational evolution from their initial particle distributions representing initial underdensities (top left) and overdensities (top right), respectively.
    Note that we illustrate only the density troughs and voids for the Reference simulation, and their corresponding density peaks and clusters for the Mirror simulation, for simplicity.
    }
    \label{fig:schematic}
\end{figure*}

In Figure~\ref{fig:schematic}, we illustrate how CCVs are identified using the Reference and Mirror simulations. We note that the cluster-scale dark matter halos in the Mirror simulation become large voids in the Reference simulation. Therefore, the dark matter particles belonging to each cluster-scale halo in one simulation can be used to define the void region in the other one. Thus, each cluster in the Mirror simulation uniquely defines a CCV in the Reference simulation.

We generate a catalog containing the position, effective radius, mean and central densities of CCVs in the Reference simulation that correspond to $422,818$ halos with halo mass $M_{\rm h}\ge10^{13}\Msun$ at $z=0$ in the Mirror simulation. The volume of a CCV is defined as the region occupied by the void member particles in the Reference simulation. A CCV associated with a more massive dark matter halo tends to have a larger volume \citep{shim+21}.

\subsection{CCVs from dark matter density fields}\label{subsec:CCVfromDM}
In this subsection, we describe a method developed by \citet{shim+21} for finding the CCVs in the Reference simulation without resorting to the Mirror simulation. First, we find the most underdense regions in the smoothed density field below a certain threshold density, and then select those whose volume is larger than a certain fraction of the typical void volume. The expected typical or mean void volume is directly related with the corresponding cluster mass. Namely, a choice of cluster mass determines the corresponding CCV size, which is used to select the void core regions. Our method is based on the finding that a larger CCV tends to develop a larger \emph{void core}, the central region with density below the specific threshold value.

There are three parameters used in this method: Gaussian smoothing scale ($R_{\rm s}$), core density threshold ($\delta_{\rm c}$), and core volume fraction ($f_{\rm c}$) with respect to the CCV volume at that mass scale. The optimal values of these parameters change depending on the mean void size or corresponding cluster mass. The matter density field is smoothed with a Gaussian filter over $R_{\rm s}$, and the void core regions with overdensity below $\delta_{\rm c}$ are found. Among them those with volume larger than $f_{\rm c}$ times the typical volume of the CCVs under interest are selected.
Because void cores only cover the innermost volume of CCVs, we then grow the cores by repeatedly attaching neighboring higher density volume elements. This growing process stops when the total volume of the recovered voids reaches that of the CCVs of the interested mass scale in the Reference simulation.

\section{Identifying CCVs from galaxy density fields}\label{sec:result}
We test whether or not the CCV identification method originally developed for dark matter density fields could be also applied to galaxy density fields. The success of the extension of the prescription depends on how accurately the galaxy density field follows the underlying dark matter density field, especially in underdense regions. To be more specific, the rank order of pixel density should be preserved; a region with a higher dark matter density should have a higher galaxy density value if we aim to identify the same voids in the galaxy and dark matter density fields using this method. Any reversal in the rank order will disturb the correspondence. Thus, we first examine the relation between galaxy and dark matter density fields in Section~\ref{sub:compare} and determine the optimal values of free parameters for galaxy density fields in Section~\ref{sub:optimal}.

\subsection{Dark matter versus Galaxy density fields}\label{sub:compare}
\begin{figure}
	\includegraphics[width=\columnwidth]{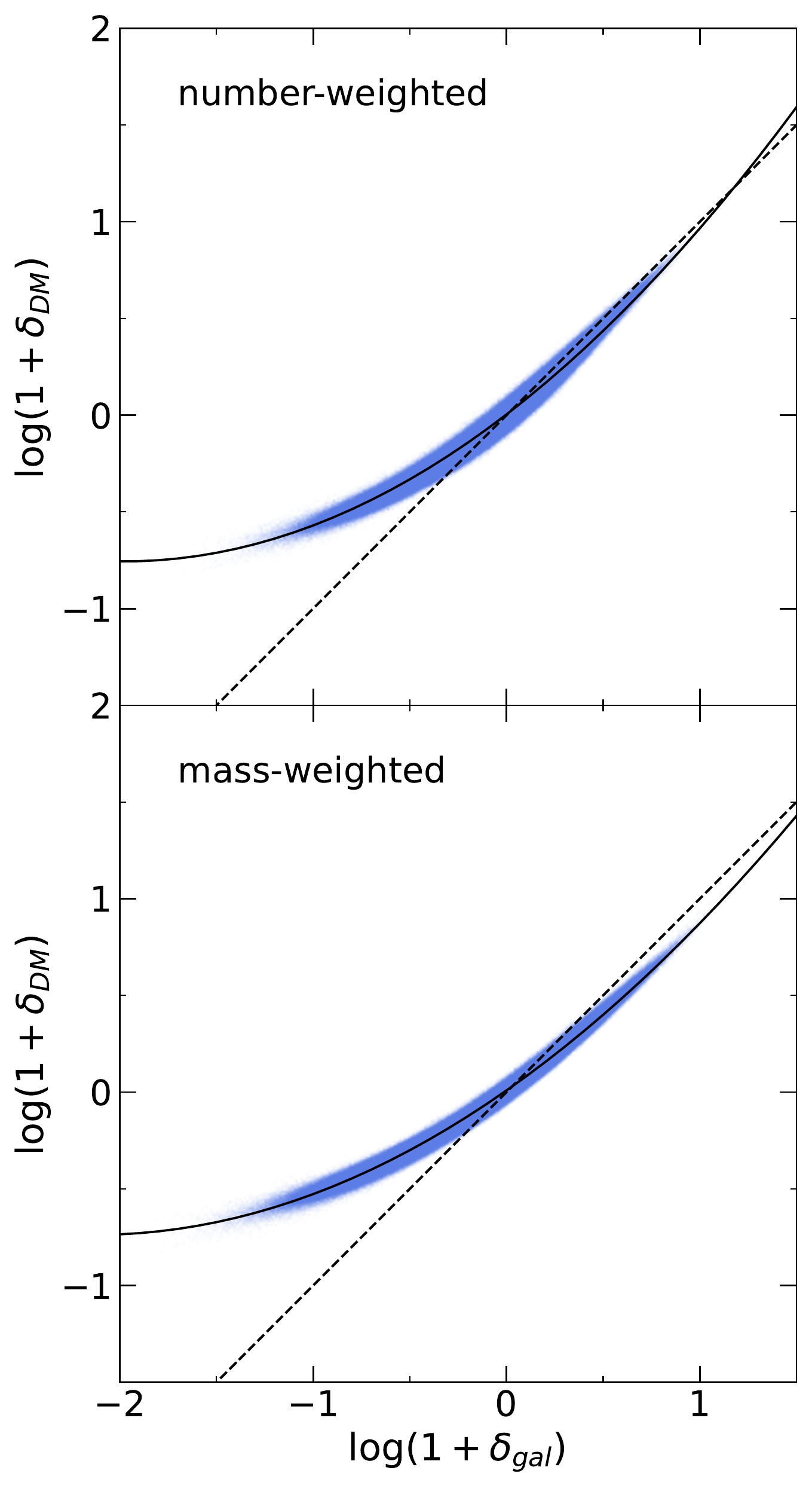}
    \caption{Relation between dark matter and galaxy overdensities. The upper (lower) panel shows the case for the number- (mass-)weighted galaxy density field.
    The Gaussian smoothing scale applied is $R_{s}=7.3\mpcph$. Solid lines are the second-order polynomial fits to the relations \citep{jee+12}, whereas dash lines represent the identity relation, $\delta_{\rm DM}=\delta_{\rm gal}$.}
    \label{fig:bias}
\end{figure}

Before we develop a void finding algorithm for a galaxy sample, we first compare between number- and mass-weighted galaxy density fields to study which one has tighter relation with the underlying matter distribution. The weighted density is calculated at the center of each pixel with a volume $(2\mpcph)^{3}$ using the cloud-in-cell (CIC) assignment scheme. We then smooth these density fields with Gaussian smoothing kernel. In Figure~\ref{fig:bias}, we compare between smoothed dark matter and galaxy density fields. As can be seen, the overdensities of both galaxy density fields are enhanced compared to that of the dark matter density field, which reflects galaxies being the biased tracers of the underlying matter field \citep{kaiser84,bardeen+86,desjacques+18}. Interestingly, the non-linear relation between dark matter and galaxy density fields is well modeled by the second-order polynomials of logarithmic densities as discovered in \citet{jee+12}. Note that this relation reduces to a linear bias model when $|\delta|\ll 1$.

We find that the mass-weighted galaxy density is more tightly correlated with the dark matter density than the number-weighted case. The standard deviation of the logarithmic dark matter overdensity for a given logarithmic galaxy overdensity bin is on average 10\% smaller in the mass-weighted galaxy field than in the number-weighted one. This is consistent with the previous results showing that mass weighting reduces the scatter between dark matter and halo density field \citep{park+07,seljak+09, park+10}. However, the standard deviation for the mass-weighted galaxy density field becomes comparable to that for the number-weighted case at an extremely low-density range. For this reason, we consider both the number- and mass-weighted galaxy density fields even though the scatter is overall smaller in the mass-weighted case. The scatter in the relation between the galaxy and dark matter fields reflects the stochastic nature of bias \citep{pen98,dekel&lahav99,matsubara99} and the shot noise effect on density measurement.
Because of these two effects, the rank order of pixel values in the density array changes when we move from the matter density field to galaxy density field. 

\begin{figure}
	\includegraphics[width=\columnwidth]{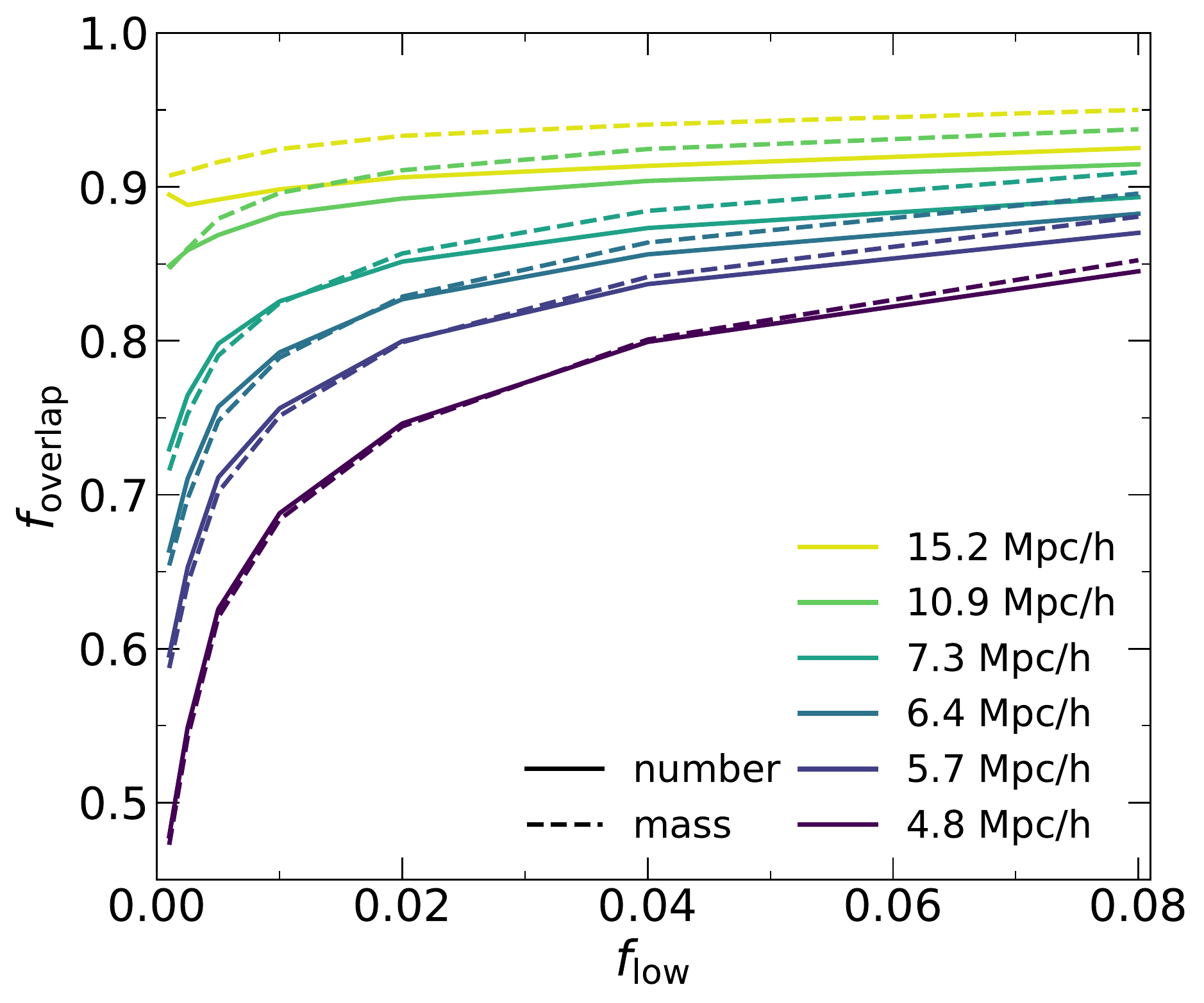}
    \caption{Volume overlap fraction ($f_{\rm overlap}$) of underdense regions between dark matter and galaxy density fields as a function of the volume fraction ($f_{\rm low}$) of the low-density regions. The galaxy density fields are constructed using the dense galaxy sample. Solid (dashed) lines represent number (mass)-weighted galaxy density fields. Different colors indicate various Gaussian smoothing scales.}
    \label{fig:matchfrac}
\end{figure}

In Figure~\ref{fig:matchfrac}, we show how much the degree of randomization in the rank order depends on the density weighting scheme and smoothing scale. We compute the volume overlap fraction ($f_{\rm overlap}$) of the lowest-density regions between galaxy and dark matter density fields as
\begin{equation}
    f_{\rm overlap} = \frac{V_{\rm overlap}}{V_{\rm low}},
\end{equation}
where $V_{\rm low}$ is the volume with density lower than the input threshold and $V_{\rm overlap}$ is the overlap volume between those lowest-density regions in the two density fields.
And the fraction of volume occupied by the lowest-density regions is given as
\begin{equation}
    f_{\rm low}=\frac{V_{\rm low}}{V_{\rm sim}},
\end{equation}
where $V_{\rm sim}$ is the entire simulation volume. For example,  $f_{\rm overlap}=0$ at $f_{\rm low}=0.02$ means that the most underdense $2\%$ volumes in galaxy and dark matter density field have no overlap between them.

We find the decreasing trend of the volume overlap fraction toward a lower $f_{\rm low}$, which becomes steeper for a smaller smoothing scale. This implies that the order shuffling between two density fields is severe at the center of voids, or in the most underdense regions. It is also shown that the overlap fraction is higher for a larger smoothing scale implying that the shuffling becomes less significant when galaxy density fields are smoothed on relatively larger smoothing scales. However, the density order at extreme low-density range is still changed by $\sim10$ per cent for the largest smoothing scale.

\subsection{Optimal parameter values for CCV Identification}\label{sub:optimal}
We need to fine-tune the free parameters for CCV identification in the galaxy density field. This is because galaxy density field is non-linearly biased with respect to the underlying dark matter density field, and hence the fine-tuned parameters obtained for the dark matter field can not be applied to the galaxy density field without adjustment. We thus follow the approach described in \citet{shim+21}. The optimal values of the free parameters are determined so that completeness and reliability are maximized.
We calculate the completeness and reliability of void finding given by
\begin{equation}
    \mathcal{C} \equiv N_{\rm s}/N_{\rm v},
\end{equation}
and 
\begin{equation}
    \mathcal{R} \equiv N_{\rm s}/N_{\rm c},
\end{equation}
respectively. Here, $N_{\rm v}$, $N_{\rm c}$, and $N_{\rm s}$ represent the number of the CCVs of a target mass scale, all the identified void cores, and  successfully-reproduced CCVs, respectively. From here on, the mass scale of a CCV refers to the corresponding cluster mass.

The completeness represents the successfully recovered fraction among the CCVs with the corresponding cluster mass above $M_{\rm min}$. On the other hand, the reliability is the fraction of recovered voids with the corresponding $M\geq 0.8M_{\rm min}$ (to allow for a buffer in mass) among all the identified voids by this method. Here, it is counted as a successful detection when the nearest CCV from the identified void core satisfies the mass criterion or when the most massive CCV within $1.5r_{\rm c}$ from the void core satisfies the mass criterion. Here, $r_{\rm c}$ is the effective radius of the void core and is measured from the void core volume $V_{\rm c}$ using the following relation
\begin{equation}
    r_{\rm c} = \left( \frac{3V_{\rm c}}{4\pi} \right)^\frac{1}{3}.
\end{equation}

\begin{table*}
	\centering
	\caption{Optimal values of the free parameters for the CCV identification from galaxy density fields constructed using the dense and sparse galaxy samples. The upper (lower) half of the table corresponds to the number- (mass-)weighted galaxy density fields. Listed in columns are the mean galaxy number density $n_{\rm gal}$, minimum target mass scale of a CCV $M_{\rm min}$, effective radius of the CCV $R_{\rm v}$, smoothing scale $R_{\rm s}$, core density threshold $\delta_{\rm c}$, upper density threshold $\delta_{\rm upper}$ when core growing stops, and minimum core fraction $f_{\rm c}$.}
	\label{tab:params}
	\begin{tabular}{cccccccc} 
		\hline\hline
		& $\bar{\it n}_{\rm gal}$ & $M_{\rm min}$ & $R_{\rm v}$ & $R_{\rm s}$ & $\delta_{\rm c}$ & $\delta_{\rm upper}$ & $f_{\rm c}$\\ 
		& [(Mpc/$\it h)^{-3}$] & [$10^{14}{\rm M}_{\odot}/h$] & [Mpc/$h$] & [R$_{\rm v}$] & & & \\
		\hline

		number-& $1.0\times10^{-2}$ & $3$ & 14.5 & 1/4 & -0.981 & -0.785 & 0.02\\
		weighted& & $5$ & 17.2 & 1/3 & -0.915 & -0.738 & 0.02\\
		& & $7$ & 19.3 & 1/3 & -0.923 & -0.725 & 0.01\\
		& & $10$ & 21.7 & 1/3 & -0.890 & -0.727 & 0.05\\
		\hline
		
		& $4.4\times10^{-3}$ & $3$ & 14.5 & 1/4 & -0.995 & -0.853 & 0.01\\
		& & $5$ & 17.2 & 1/4 & -0.981 & -0.858 & 0.05\\
		& & $7$ & 19.3 & 1/4 & -0.967 & -0.861 & 0.10\\
		& & $10$ & 21.7 & 1/2 & -0.835 & -0.619 & 0.01\\
		\hline\hline
		
		mass-& $1.0\times10^{-2}$ & $3$ & 14.5 & 1/4 & -0.987 & -0.835 & 0.02\\
		weighted& & $5$ & 17.2 & 1/3 & -0.928  & -0.797 & 0.03\\
		& & $7$ & 19.3 & 1/3 & -0.942  & -0.779 & 0.01\\
		& & $10$ & 21.7 & 1/3 & -0.921 & -0.771 & 0.04\\
		\hline
		
		& $4.4\times10^{-3}$ & $3$ & 14.5 & 1/4 & -0.996  & -0.894 & 0.01\\
		& & $5$ & 17.2 & 1/3 & -0.957  & -0.829 & 0.02\\
		& & $7$ & 19.3 & 1/3 & -0.950  & -0.824 & 0.03\\
		& & $10$ & 21.7 & 1/3 & -0.937  & -0.816 & 0.05\\
		\hline
	\end{tabular}
\end{table*}

We repeat calculating the detection completeness and reliability in three dimensional parameter space of $R_{\rm s}$, $\delta_{\rm c}$, and $f_{\rm c}$, and provide their optimal values in Table~\ref{tab:params}. Note that we limit the smoothing scales to be the reciprocals of integers multiplied by the effective radius of the CCVs of the target mass scale. The optimal length scales for smoothing galaxy density fields are $R_{\rm s}=R_{\rm v}/3$ in most cases, where $R_{\rm v}$ is the effective radius of a CCV of a particular mass scale. This implies that the optimized smoothing scale of a given galaxy density field is determined by the target mass scale, or equivalently the target size, of voids. Interestingly, the relation between the smoothing scale and void size is consistent with the case for the dark matter density fields \citep{shim+21}.

\begin{figure}
	\includegraphics[width=\columnwidth]{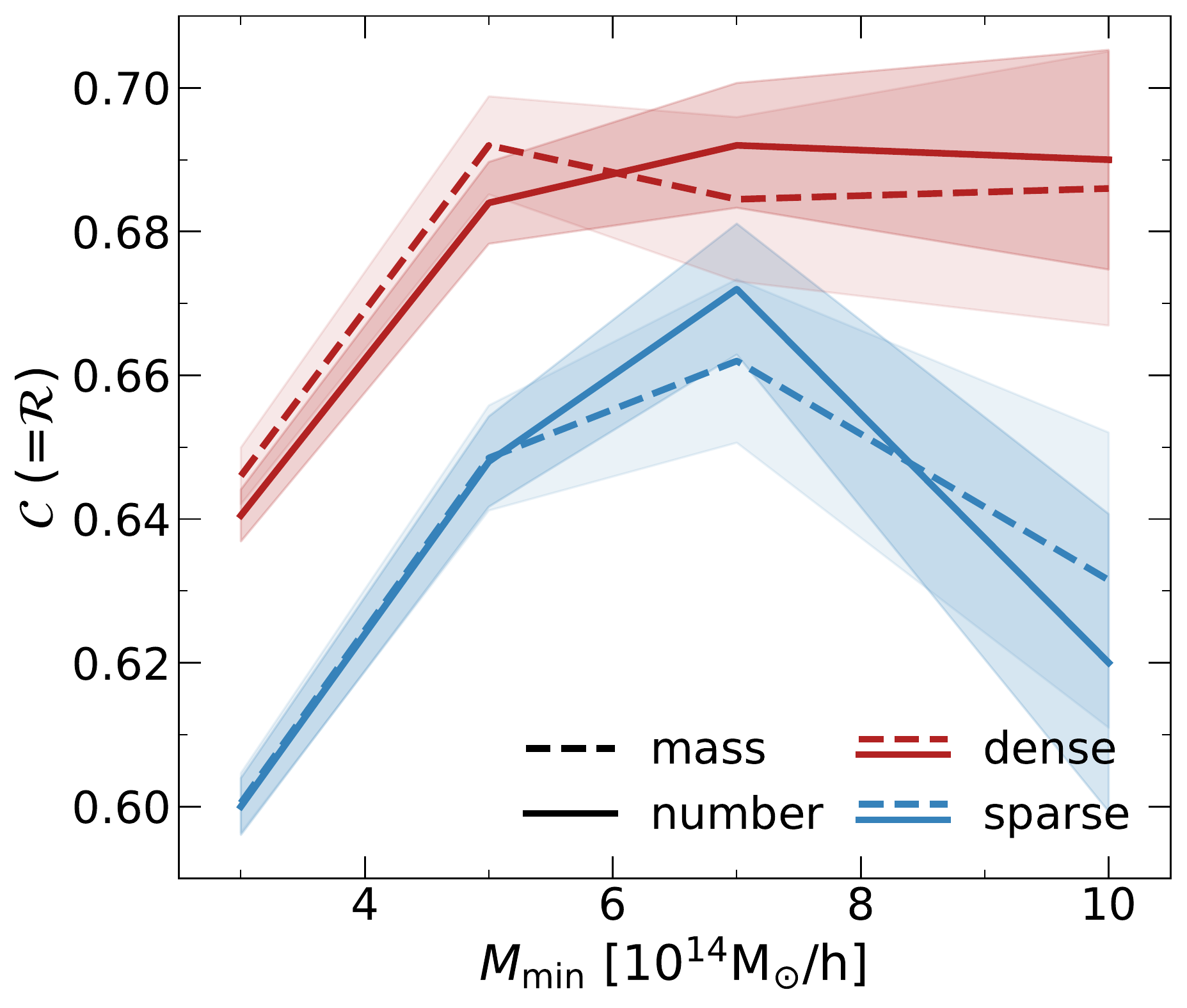}
    \caption{Completeness and reliability of identifying CCVs from the mass-weighted (dashed) and number-weighted (solid) density fields constructed using the dense (red) and sparse (blue) galaxy sample. The shaded regions are the detection rate uncertainties calculated as the standard deviations of the completeness and reliability measured from bootstrap resampled galaxy density fields. The detection rates and uncertainties are calculated when the optimal parameter values in Table~\ref{tab:params} are adopted. Note that the completeness and reliability are identical for the optimal parameter values.}
    \label{fig:CR}
\end{figure}

In Figure~\ref{fig:CR}, we show the detection completeness and reliability for different minimum target mass scales of CCVs. It is shown that the detection completeness, and equivalently the reliability, does not depend much on the density-weighting scheme. On the other hand, the mean galaxy number density significantly affects the detection quality. This is just because of the unavoidable nature of the cosmic voids whose identification is sensitive to the tracer number density. For the sparse sample, the detection rates increase from $\mathcal{C}=\mathcal{R}=0.60$ to $0.66$ with the void mass scale before $M_{\rm min}\simeq 7\times 10^{14}\msolph$, and then they drop to $0.62$ at $M_{\rm min}=10^{15}\msolph$. When using the dense sample, the detection rates are enhanced approximately by $2-7\%$. This is because the dense sample includes low mass galaxies that are more likely to be in voids than in other environments \citep{alonso+15}. Thus, increasing sample density decreases the shot noise effect, and hence the density-order shuffling in underdense regions. The detection rate enhancement is most noticeable at the largest mass scale, $M_{\rm min}=10^{15}\msolph$. Because the halo abundance in a large void is lower than in a small void \citep{patiri+06}, increasing the sample density by adding lower-mass galaxies has more significant effect on decreasing the shot-noise for larger voids. Consequently, the density-order shuffling diminishes, leading to the rapid increase in the detection rate at the largest mass scale. In line with this interpretation, we find that the uncertainty of the detection rate, which increases with the void mass scale, are the largest for the largest scale voids. We compute the uncertainty as the standard deviation of the detection rate measured from 1000 realizations of bootstrap resampled galaxy density fields. More importantly, increasing sample density reduces the uncertainties of the detection rates approximately by $26\%$ and $7\%$ at the largest void mass scale, whereas they only decreased by $7.6\%$ and $3.7\%$ on average on smaller mass scales for the number- and mass-weighted cases, respectively. Thus, the benefit of increasing galaxy sample density is most significant in the largest CCV detection.

\begin{figure*}
	\includegraphics[width=2\columnwidth]{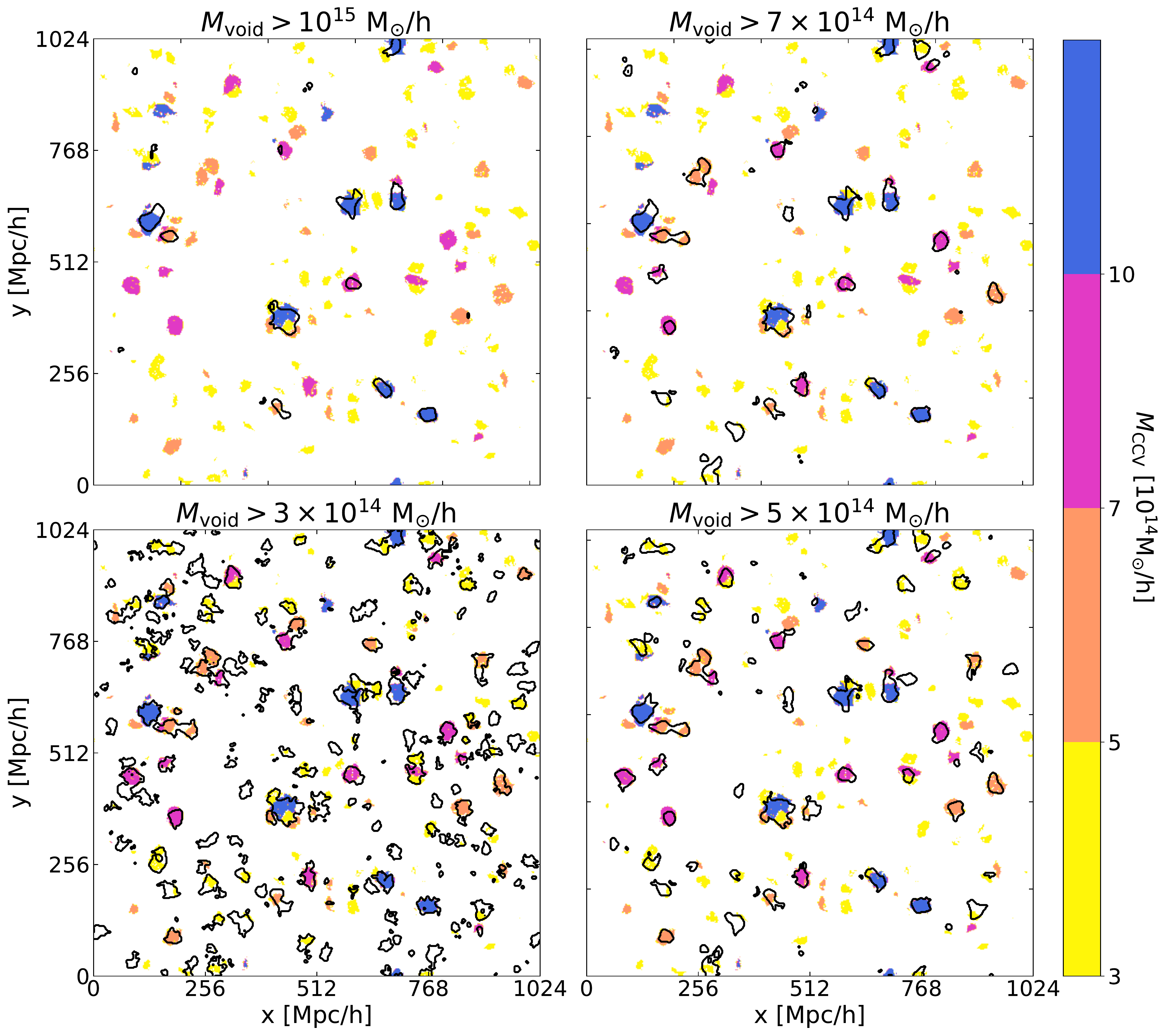}
    \caption{Recovered voids (black contours) from the number-weighted galaxy density fields and CCVs (non-black) of various mass scales in a $2\mpcph$-thick slice. The optimal parameter values listed in Table~\ref{tab:params} are adopted. All panels show the identical field. Recovered voids corresponding to galaxy cluster mass scale above $1\times10^{15},\ 7\times 10^{14},\ 5\times10^{14}$ and $3\times10^{14} \msolph$ are shown in clockwise direction from the upper-left panel to bottom-left.}
    \label{fig:grownseed}
\end{figure*}

Finally, we show in Figure~\ref{fig:grownseed} the spatial correspondence between the recovered voids (black contours) using this method and the CCVs (colored regions) defined under the void-cluster correspondence theory. The voids are recovered from the number-weighted galaxy density fields constructed using the dense galaxy sample. It is shown that the recovered void regions well overlap with the CCVs of the target mass scales. A recovered void typically corresponds to a single CCV unless the corresponding CCV has neighboring CCVs. This nearly one-to-one correspondence between a recovered void and a CCV becomes relatively weaker for a smaller minimum target void mass. Thus, the trend of a recovered void encompassing multiple CCVs is most noticeable for $M_{\rm void}>3\times10^{14}\msolph$. However, in principle, by comparing recovered void regions for two different target mass scales one can decompose such void complexes into several smaller voids that respectively correspond to single CCVs.

\section{Summary and Discussion}\label{sec:discussion and summary}

We identify cosmic voids from galaxy density fields under the void-cluster correspondence theory. The correspondence between voids and clusters of the same mass scale is established using paired cosmological simulations with sign-opposite initial overdensity fields. Cluster-counterpart voids (CCVs) in one simulation are defined as the regions occupied by the same dark matter particles that form cluster-scale halos in its inverted simulation.

Extending the CCV-identification method \citep{shim+21} developed for matter density fields, we find the optimal values of the free parameters for the void identification in galaxy density fields; Gaussian smoothing scale, density threshold, and core volume fraction. The optimal length scale for smoothing galaxy density fields is determined by the target mass scale of voids, which is related to the target void radius as $R_{\rm s} = R_{\rm v}/3$. The optimal parameter values for the number- and mass- weighted density fields constructed using the sparse and dense galaxy samples are listed in Table~\ref{tab:params}. Using the dense galaxy sample, we can achieve about $64-69\%$ of CCV detection completeness and reliability from the number-weighted galaxy density fields. The detection rates decrease by $\sim2-7\%$ when using the sparse galaxy sample due to increasing effects of shot-noise on density.

Our result demonstrates that we can identify CCVs from galaxy density fields in real space. In actual galaxy surveys, on the other hand, one needs to consider the redshift-space distortion (RSD) when constructing density fields from the observed galaxy distribution in redshift space. Directly identifying CCVs in redshift space without considering the RSD effect will be less reliable than in real space. This is because the RSD further introduces the density-rank order randomization in addition to those induced by the shot-noise and stochasticity of bias. Thus, it is better to adopt reconstruction techniques and identify CCVs in real space than in redshift space. The fingers of God
\citep{jackson72} due to the internal motion of galaxies within clusters can be effectively removed by forcing the line-of-sight elongation of clusters equal to their size perpendicular to the line-of-sight \citep{tegmark+04,park+12,tully15, hwang+16}. On larger scales, the Kaiser effect \citep{kaiser87} due to coherent flow induced by large-scale over/under-densities can be corrected using Zel'dovich approximation \citep{zeldovich70} or second order Lagrangian perturbation theory \citep{scoccimarro98}. Using these approaches, one can place galaxies back to their real space positions by subtracting the displacement due to their peculiar motions \citep{wang+09, kitaura+12, bos+19}.

Identifying CCVs from observational data could help us perform more precise cosmological and astrophysical analysis using voids.
Because CCVs have universal mass density distribution \citep{shim+21} and common central and average densities \citep{pontzen+16,stopyra+21,desmond+22}, they represent relatively more homogeneous void population than arbitrary underdense regions extracted from density fields. Consequently, when using CCVs it is possible to decrease the uncertainties arising due to the scatter in void properties. This will be the subject of a future study. One of the possible applications of CCVs in the cosmological context is to measure the linear growth rate using CCVs. In addition to the homogenous properties of the CCVs, their internal dynamics remain
near linear \citep{stopyra+21}, and the linear bias relation holds at around CCVs \citep{pollina+17}.
Because the linearity in and around the CCV environment allows a simple linear modeling for void-galaxy cross-correlation accurate measurements of linear growth rate using voids \citep{achitouv19,hamaus+22,woodfinden+22} can be made using CCVs. We leave such cosmological analysis using CCVs from galaxy surveys to future work.

\section*{Acknowledgements}
We thank an anonymous referee for helpful comments that helped improve the original manuscript. J.S. and C.B.P. were supported by KIAS Individual Grant PG071202 and PG016903 at Korea Institute for Advanced Study, respectively. J.K. was supported by a KIAS Individual Grant (KG039603) via the Center for Advanced Computation at Korea Institute for Advanced Study. S.E.H. was supported by the project 우주거대구조를 이용한 암흑우주연구 (“Understanding the Dark Universe Using Large Scale Structure of the Universe”), funded by the Ministry of Science. The computing resources were kindly provided by the Center for Advanced Computation at Korea Institute for Advanced Study.

\bibliography{reference}

\begin{thebibliography}{}
\expandafter\ifx\csname natexlab\endcsname\relax\def\natexlab#1{#1}\fi
\providecommand{\url}[1]{\href{#1}{#1}}

\bibitem[{{Achitouv}(2017)}]{achitouv17}
{Achitouv}, I. 2017, \prd, 96, 083506

\bibitem[{{Achitouv}(2019)}]{achitouv19}
---. 2019, \prd, 100, 123513

\bibitem[{{Achitouv} {et~al.}(2015){Achitouv}, {Neyrinck}, \&
  {Paranjape}}]{achitouv+15}
{Achitouv}, I., {Neyrinck}, M., \& {Paranjape}, A. 2015, \mnras, 451, 3964

\bibitem[{{Alonso} {et~al.}(2015){Alonso}, {Eardley}, \& {Peacock}}]{alonso+15}
{Alonso}, D., {Eardley}, E., \& {Peacock}, J.~A. 2015, \mnras, 447, 2683

\bibitem[{{Bardeen} {et~al.}(1986){Bardeen}, {Bond}, {Kaiser}, \&
  {Szalay}}]{bardeen+86}
{Bardeen}, J.~M., {Bond}, J.~R., {Kaiser}, N., \& {Szalay}, A.~S. 1986, \apj,
  304, 15

\bibitem[{{Bertschinger}(1985)}]{bertschinger85}
{Bertschinger}, E. 1985, \apjs, 58, 1

\bibitem[{{Beygu} {et~al.}(2013){Beygu}, {Kreckel}, {van de Weygaert}, {van der
  Hulst}, \& {van Gorkom}}]{beygu+13}
{Beygu}, B., {Kreckel}, K., {van de Weygaert}, R., {van der Hulst}, J.~M., \&
  {van Gorkom}, J.~H. 2013, \aj, 145, 120

\bibitem[{{Bos} {et~al.}(2019){Bos}, {Kitaura}, \& {van de Weygaert}}]{bos+19}
{Bos}, E.~G.~P., {Kitaura}, F.-S., \& {van de Weygaert}, R. 2019, \mnras, 488,
  2573

\bibitem[{{Cai} {et~al.}(2015){Cai}, {Padilla}, \& {Li}}]{cai+15}
{Cai}, Y.-C., {Padilla}, N., \& {Li}, B. 2015, \mnras, 451, 1036

\bibitem[{{Ceccarelli} {et~al.}(2022){Ceccarelli}, {Duplancic}, \& {Garcia
  Lambas}}]{ceccarelli+22}
{Ceccarelli}, L., {Duplancic}, F., \& {Garcia Lambas}, D. 2022, \mnras, 509,
  1805

\bibitem[{{Ceccarelli} {et~al.}(2008){Ceccarelli}, {Padilla}, \&
  {Lambas}}]{ceccarelli+08}
{Ceccarelli}, L., {Padilla}, N., \& {Lambas}, D.~G. 2008, \mnras, 390, L9

\bibitem[{{Chan} {et~al.}(2014){Chan}, {Hamaus}, \& {Desjacques}}]{chan+14}
{Chan}, K.~C., {Hamaus}, N., \& {Desjacques}, V. 2014, \prd, 90, 103521

\bibitem[{{Choi} {et~al.}(2010){Choi}, {Park}, {Kim}, {Gott}, {Weinberg},
  {Vogeley}, {Kim}, \& {SDSS Collaboration}}]{choi+10best}
{Choi}, Y.-Y., {Park}, C., {Kim}, J., {et~al.} 2010, \apjs, 190, 181

\bibitem[{{Choi} {et~al.}(2007){Choi}, {Park}, \& {Vogeley}}]{choi+07}
{Choi}, Y.-Y., {Park}, C., \& {Vogeley}, M.~S. 2007, \apj, 658, 884

\bibitem[{{Colberg} {et~al.}(2005){Colberg}, {Sheth}, {Diaferio}, {Gao}, \&
  {Yoshida}}]{colberg+05}
{Colberg}, J.~M., {Sheth}, R.~K., {Diaferio}, A., {Gao}, L., \& {Yoshida}, N.
  2005, \mnras, 360, 216

\bibitem[{{de Lapparent} {et~al.}(1986){de Lapparent}, {Geller}, \&
  {Huchra}}]{lapparent+86}
{de Lapparent}, V., {Geller}, M.~J., \& {Huchra}, J.~P. 1986, \apjl, 302, L1

\bibitem[{{De Lucia} {et~al.}(2004){De Lucia}, {Kauffmann}, \&
  {White}}]{delucia+04}
{De Lucia}, G., {Kauffmann}, G., \& {White}, S. D.~M. 2004, \mnras, 349, 1101

\bibitem[{{Dekel} \& {Lahav}(1999)}]{dekel&lahav99}
{Dekel}, A., \& {Lahav}, O. 1999, \apj, 520, 24

\bibitem[{{DESI Collaboration} {et~al.}(2016){DESI Collaboration}, {Aghamousa},
  {Aguilar}, {Ahlen}, {Alam}, {Allen}, {Allende Prieto}, {Annis}, {Bailey},
  {Balland}, {Ballester}, {Baltay}, {Beaufore}, {Bebek}, {Beers}, {Bell},
  {Bernal}, {Besuner}, {Beutler}, {Blake}, {Bleuler}, {Blomqvist}, {Blum},
  {Bolton}, {Briceno}, {Brooks}, {Brownstein}, {Buckley-Geer}, {Burden},
  {Burtin}, {Busca}, {Cahn}, {Cai}, {Cardiel-Sas}, {Carlberg}, {Carton},
  {Casas}, {Castander}, {Cervantes-Cota}, {Claybaugh}, {Close}, {Coker},
  {Cole}, {Comparat}, {Cooper}, {Cousinou}, {Crocce}, {Cuby}, {Cunningham},
  {Davis}, {Dawson}, {de la Macorra}, {De Vicente}, {Delubac}, {Derwent},
  {Dey}, {Dhungana}, {Ding}, {Doel}, {Duan}, {Ealet}, {Edelstein},
  {Eftekharzadeh}, {Eisenstein}, {Elliott}, {Escoffier}, {Evatt}, {Fagrelius},
  {Fan}, {Fanning}, {Farahi}, {Farihi}, {Favole}, {Feng}, {Fernandez},
  {Findlay}, {Finkbeiner}, {Fitzpatrick}, {Flaugher}, {Flender}, {Font-Ribera},
  {Forero-Romero}, {Fosalba}, {Frenk}, {Fumagalli}, {Gaensicke}, {Gallo},
  {Garcia-Bellido}, {Gaztanaga}, {Pietro Gentile Fusillo}, {Gerard},
  {Gershkovich}, {Giannantonio}, {Gillet}, {Gonzalez-de-Rivera},
  {Gonzalez-Perez}, {Gott}, {Graur}, {Gutierrez}, {Guy}, {Habib}, {Heetderks},
  {Heetderks}, {Heitmann}, {Hellwing}, {Herrera}, {Ho}, {Holland}, {Honscheid},
  {Huff}, {Hutchinson}, {Huterer}, {Hwang}, {Illa Laguna}, {Ishikawa},
  {Jacobs}, {Jeffrey}, {Jelinsky}, {Jennings}, {Jiang}, {Jimenez}, {Johnson},
  {Joyce}, {Jullo}, {Juneau}, {Kama}, {Karcher}, {Karkar}, {Kehoe}, {Kennamer},
  {Kent}, {Kilbinger}, {Kim}, {Kirkby}, {Kisner}, {Kitanidis}, {Kneib},
  {Koposov}, {Kovacs}, {Koyama}, {Kremin}, {Kron}, {Kronig}, {Kueter-Young},
  {Lacey}, {Lafever}, {Lahav}, {Lambert}, {Lampton}, {Landriau}, {Lang},
  {Lauer}, {Le Goff}, {Le Guillou}, {Le Van Suu}, {Lee}, {Lee}, {Leitner},
  {Lesser}, {Levi}, {L'Huillier}, {Li}, {Liang}, {Lin}, {Linder}, {Loebman},
  {Luki{\'c}}, {Ma}, {MacCrann}, {Magneville}, {Makarem}, {Manera}, {Manser},
  {Marshall}, {Martini}, {Massey}, {Matheson}, {McCauley}, {McDonald},
  {McGreer}, {Meisner}, {Metcalfe}, {Miller}, {Miquel}, {Moustakas}, {Myers},
  {Naik}, {Newman}, {Nichol}, {Nicola}, {Nicolati da Costa}, {Nie}, {Niz},
  {Norberg}, {Nord}, {Norman}, {Nugent}, {O'Brien}, {Oh}, {Olsen}, {Padilla},
  {Padmanabhan}, {Padmanabhan}, {Palanque-Delabrouille}, {Palmese},
  {Pappalardo}, {P{\^a}ris}, {Park}, {Patej}, {Peacock}, {Peiris}, {Peng},
  {Percival}, {Perruchot}, {Pieri}, {Pogge}, {Pollack}, {Poppett}, {Prada},
  {Prakash}, {Probst}, {Rabinowitz}, {Raichoor}, {Ree}, {Refregier}, {Regal},
  {Reid}, {Reil}, {Rezaie}, {Rockosi}, {Roe}, {Ronayette}, {Roodman}, {Ross},
  {Ross}, {Rossi}, {Rozo}, {Ruhlmann-Kleider}, {Rykoff}, {Sabiu}, {Samushia},
  {Sanchez}, {Sanchez}, {Schlegel}, {Schneider}, {Schubnell}, {Secroun},
  {Seljak}, {Seo}, {Serrano}, {Shafieloo}, {Shan}, {Sharples}, {Sholl},
  {Shourt}, {Silber}, {Silva}, {Sirk}, {Slosar}, {Smith}, {Smoot}, {Som},
  {Song}, {Sprayberry}, {Staten}, {Stefanik}, {Tarle}, {Sien Tie}, {Tinker},
  {Tojeiro}, {Valdes}, {Valenzuela}, {Valluri}, {Vargas-Magana}, {Verde},
  {Walker}, {Wang}, {Wang}, {Weaver}, {Weaverdyck}, {Wechsler}, {Weinberg},
  {White}, {Yang}, {Yeche}, {Zhang}, {Zhao}, {Zheng}, {Zhou}, {Zhou}, {Zhu},
  {Zou}, \& {Zu}}]{DESI+16}
{DESI Collaboration}, {Aghamousa}, A., {Aguilar}, J., {et~al.} 2016, arXiv
  e-prints, arXiv:1611.00036

\bibitem[{{Desjacques} {et~al.}(2018){Desjacques}, {Jeong}, \&
  {Schmidt}}]{desjacques+18}
{Desjacques}, V., {Jeong}, D., \& {Schmidt}, F. 2018, \physrep, 733, 1

\bibitem[{{Desmond} {et~al.}(2022){Desmond}, {Hutt}, {Devriendt}, \&
  {Slyz}}]{desmond+22}
{Desmond}, H., {Hutt}, M.~L., {Devriendt}, J., \& {Slyz}, A. 2022, \mnras, 511,
  L45

\bibitem[{{Dor{\'e}} {et~al.}(2014){Dor{\'e}}, {Bock}, {Ashby}, {Capak},
  {Cooray}, {de Putter}, {Eifler}, {Flagey}, {Gong}, {Habib}, {Heitmann},
  {Hirata}, {Jeong}, {Katti}, {Korngut}, {Krause}, {Lee}, {Masters},
  {Mauskopf}, {Melnick}, {Mennesson}, {Nguyen}, {{\"O}berg}, {Pullen},
  {Raccanelli}, {Smith}, {Song}, {Tolls}, {Unwin}, {Venumadhav}, {Viero},
  {Werner}, \& {Zemcov}}]{SPHEREx+14}
{Dor{\'e}}, O., {Bock}, J., {Ashby}, M., {et~al.} 2014, arXiv e-prints,
  arXiv:1412.4872

\bibitem[{{Dubinski} {et~al.}(2004){Dubinski}, {Kim}, {Park}, \&
  {Humble}}]{dubinski+04}
{Dubinski}, J., {Kim}, J., {Park}, C., \& {Humble}, R. 2004, \na, 9, 111

\bibitem[{{Dunkley} {et~al.}(2009){Dunkley}, {Komatsu}, {Nolta}, {Spergel},
  {Larson}, {Hinshaw}, {Page}, {Bennett}, {Gold}, {Jarosik}, {Weiland},
  {Halpern}, {Hill}, {Kogut}, {Limon}, {Meyer}, {Tucker}, {Wollack}, \&
  {Wright}}]{dunkley+09}
{Dunkley}, J., {Komatsu}, E., {Nolta}, M.~R., {et~al.} 2009, \apjs, 180, 306

\bibitem[{{Einasto} {et~al.}(1980){Einasto}, {Joeveer}, \& {Saar}}]{einasto+80}
{Einasto}, J., {Joeveer}, M., \& {Saar}, E. 1980, \mnras, 193, 353

\bibitem[{{El-Ad} {et~al.}(1996){El-Ad}, {Piran}, \& {da Costa}}]{el-ad96}
{El-Ad}, H., {Piran}, T., \& {da Costa}, L.~N. 1996, \apjl, 462, L13

\bibitem[{{Faltenbacher} \& {Diemand}(2006)}]{faltenbacher&diemand06}
{Faltenbacher}, A., \& {Diemand}, J. 2006, \mnras, 369, 1698

\bibitem[{{Fillmore} \& {Goldreich}(1984)}]{filmore&goldreich84}
{Fillmore}, J.~A., \& {Goldreich}, P. 1984, \apj, 281, 9

\bibitem[{{Forero-Romero} {et~al.}(2009){Forero-Romero}, {Hoffman},
  {Gottl{\"o}ber}, {Klypin}, \& {Yepes}}]{forero09}
{Forero-Romero}, J.~E., {Hoffman}, Y., {Gottl{\"o}ber}, S., {Klypin}, A., \&
  {Yepes}, G. 2009, \mnras, 396, 1815

\bibitem[{{Gott} {et~al.}(1986){Gott}, {Melott}, \& {Dickinson}}]{gott+86}
{Gott}, J.~Richard, I., {Melott}, A.~L., \& {Dickinson}, M. 1986, \apj, 306,
  341

\bibitem[{{Hahn} {et~al.}(2007){Hahn}, {Porciani}, {Carollo}, \&
  {Dekel}}]{hahn+07}
{Hahn}, O., {Porciani}, C., {Carollo}, C.~M., \& {Dekel}, A. 2007, \mnras, 375,
  489

\bibitem[{{Hamaus} {et~al.}(2022){Hamaus}, {Aubert}, {Pisani}, {Contarini},
  {Verza}, {Cousinou}, {Escoffier}, {Hawken}, {Lavaux}, {Pollina}, {Wandelt},
  {Weller}, {Bonici}, {Carbone}, {Guzzo}, {Kovacs}, {Marulli}, {Massara},
  {Moscardini}, {Ntelis}, {Percival}, {Radinovi{\'c}}, {Sahl{\'e}n}, {Sakr},
  {S{\'a}nchez}, {Winther}, {Auricchio}, {Awan}, {Bender}, {Bodendorf},
  {Bonino}, {Branchini}, {Brescia}, {Brinchmann}, {Capobianco}, {Carretero},
  {Castander}, {Castellano}, {Cavuoti}, {Cimatti}, {Cledassou}, {Congedo},
  {Conversi}, {Copin}, {Corcione}, {Cropper}, {Da Silva}, {Degaudenzi},
  {Douspis}, {Dubath}, {Duncan}, {Dupac}, {Dusini}, {Ealet}, {Ferriol},
  {Fosalba}, {Frailis}, {Franceschi}, {Franzetti}, {Fumana}, {Garilli},
  {Gillis}, {Giocoli}, {Grazian}, {Grupp}, {Haugan}, {Holmes}, {Hormuth},
  {Jahnke}, {Kermiche}, {Kiessling}, {Kilbinger}, {Kitching}, {K{\"u}mmel},
  {Kunz}, {Kurki-Suonio}, {Ligori}, {Lilje}, {Lloro}, {Maiorano}, {Marggraf},
  {Markovic}, {Massey}, {Maurogordato}, {Melchior}, {Meneghetti}, {Meylan},
  {Moresco}, {Munari}, {Niemi}, {Padilla}, {Paltani}, {Pasian}, {Pedersen},
  {Pettorino}, {Pires}, {Poncet}, {Popa}, {Pozzetti}, {Rebolo}, {Rhodes},
  {Rix}, {Roncarelli}, {Rossetti}, {Saglia}, {Schneider}, {Secroun}, {Seidel},
  {Serrano}, {Sirignano}, {Sirri}, {Starck}, {Tallada-Cresp{\'\i}},
  {Tavagnacco}, {Taylor}, {Tereno}, {Toledo-Moreo}, {Torradeflot}, {Valentijn},
  {Valenziano}, {Wang}, {Welikala}, {Zamorani}, {Zoubian}, {Andreon}, {Baldi},
  {Camera}, {Mei}, {Neissner}, \& {Romelli}}]{hamaus+22}
{Hamaus}, N., {Aubert}, M., {Pisani}, A., {et~al.} 2022, \aap, 658, A20

\bibitem[{{Hong} {et~al.}(2016){Hong}, {Park}, \& {Kim}}]{hong+16}
{Hong}, S.~E., {Park}, C., \& {Kim}, J. 2016, \apj, 823, 103

\bibitem[{{Hwang} {et~al.}(2016){Hwang}, {Geller}, {Park}, {Fabricant},
  {Kurtz}, {Rines}, {Kim}, {Diaferio}, {Zahid}, {Berlind}, {Calkins}, {Tokarz},
  \& {Moran}}]{hwang+16}
{Hwang}, H.~S., {Geller}, M.~J., {Park}, C., {et~al.} 2016, \apj, 818, 173

\bibitem[{{Jackson}(1972)}]{jackson72}
{Jackson}, J.~C. 1972, \mnras, 156, 1P

\bibitem[{{Jee} {et~al.}(2012){Jee}, {Park}, {Kim}, {Choi}, \& {Kim}}]{jee+12}
{Jee}, I., {Park}, C., {Kim}, J., {Choi}, Y.-Y., \& {Kim}, S.~S. 2012, \apj,
  753, 11

\bibitem[{{Jiang} {et~al.}(2008){Jiang}, {Jing}, {Faltenbacher}, {Lin}, \&
  {Li}}]{jiang+08}
{Jiang}, C.~Y., {Jing}, Y.~P., {Faltenbacher}, A., {Lin}, W.~P., \& {Li}, C.
  2008, \apj, 675, 1095

\bibitem[{{Joeveer} \& {Einasto}(1978)}]{joeveer&einasto78}
{Joeveer}, M., \& {Einasto}, J. 1978, in Large Scale Structures in the
  Universe, ed. M.~S. {Longair} \& J.~{Einasto}, Vol.~79, 241

\bibitem[{{Kaiser}(1984)}]{kaiser84}
{Kaiser}, N. 1984, \apjl, 284, L9

\bibitem[{{Kaiser}(1987)}]{kaiser87}
---. 1987, \mnras, 227, 1

\bibitem[{{Kauffmann} \& {Fairall}(1991)}]{kauffmann91}
{Kauffmann}, G., \& {Fairall}, A.~P. 1991, \mnras, 248, 313

\bibitem[{{Kim} \& {Park}(1998)}]{kim&park98}
{Kim}, M., \& {Park}, C. 1998, Journal of Korean Astronomical Society, 31, 109

\bibitem[{{Kitaura} {et~al.}(2012){Kitaura}, {Angulo}, {Hoffman}, \&
  {Gottl{\"o}ber}}]{kitaura+12}
{Kitaura}, F.-S., {Angulo}, R.~E., {Hoffman}, Y., \& {Gottl{\"o}ber}, S. 2012,
  \mnras, 425, 2422

\bibitem[{{Kreckel} {et~al.}(2011){Kreckel}, {Platen}, {Arag{\'o}n-Calvo}, {van
  Gorkom}, {van de Weygaert}, {van der Hulst}, {Kova{\v{c}}}, {Yip}, \&
  {Peebles}}]{kreckel+11}
{Kreckel}, K., {Platen}, E., {Arag{\'o}n-Calvo}, M.~A., {et~al.} 2011, \aj,
  141, 4

\bibitem[{{Lavaux} \& {Wandelt}(2010)}]{lavaux&wandelt10}
{Lavaux}, G., \& {Wandelt}, B.~D. 2010, \mnras, 403, 1392

\bibitem[{{Lee} \& {Park}(2009)}]{lee&park09}
{Lee}, J., \& {Park}, D. 2009, \apjl, 696, L10

\bibitem[{{Li}(2011)}]{li11}
{Li}, B. 2011, \mnras, 411, 2615

\bibitem[{{Li} {et~al.}(2012){Li}, {Zhao}, \& {Koyama}}]{li+12}
{Li}, B., {Zhao}, G.-B., \& {Koyama}, K. 2012, \mnras, 421, 3481

\bibitem[{{Matsubara}(1999)}]{matsubara99}
{Matsubara}, T. 1999, \apj, 525, 543

\bibitem[{{Nadathur} \& {Hotchkiss}(2015)}]{nadathur&hotchkiss15}
{Nadathur}, S., \& {Hotchkiss}, S. 2015, \mnras, 454, 2228

\bibitem[{{Neyrinck}(2008)}]{neyrinck08}
{Neyrinck}, M.~C. 2008, \mnras, 386, 2101

\bibitem[{{Nusser} {et~al.}(2005){Nusser}, {Gubser}, \& {Peebles}}]{nusser+05}
{Nusser}, A., {Gubser}, S.~S., \& {Peebles}, P.~J. 2005, \prd, 71, 083505

\bibitem[{{Park} {et~al.}(2012){Park}, {Choi}, {Kim}, {Gott}, {Kim}, \&
  {Kim}}]{park+12}
{Park}, C., {Choi}, Y.-Y., {Kim}, J., {et~al.} 2012, \apjl, 759, L7

\bibitem[{{Park} {et~al.}(2007){Park}, {Choi}, {Vogeley}, {Gott}, {Blanton}, \&
  {SDSS Collaboration}}]{park+07}
{Park}, C., {Choi}, Y.-Y., {Vogeley}, M.~S., {et~al.} 2007, \apj, 658, 898

\bibitem[{{Park} \& {Lee}(2009)}]{park&lee09}
{Park}, D., \& {Lee}, J. 2009, \mnras, 400, 1105

\bibitem[{{Park} {et~al.}(2010){Park}, {Kim}, \& {Park}}]{park+10}
{Park}, H., {Kim}, J., \& {Park}, C. 2010, \apj, 714, 207

\bibitem[{{Park} {et~al.}(2019){Park}, {Park}, {Sabiu}, {Li}, {Hong}, {Kim},
  {Tonegawa}, \& {Zheng}}]{park+19}
{Park}, H., {Park}, C., {Sabiu}, C.~G., {et~al.} 2019, \apj, 881, 146

\bibitem[{{Patiri} {et~al.}(2006{\natexlab{a}}){Patiri}, {Betancort-Rijo}, \&
  {Prada}}]{patiri+06}
{Patiri}, S.~G., {Betancort-Rijo}, J., \& {Prada}, F. 2006{\natexlab{a}},
  \mnras, 368, 1132

\bibitem[{{Patiri} {et~al.}(2006{\natexlab{b}}){Patiri}, {Betancort-Rijo},
  {Prada}, {Klypin}, \& {Gottl{\"o}ber}}]{patiri06}
{Patiri}, S.~G., {Betancort-Rijo}, J.~E., {Prada}, F., {Klypin}, A., \&
  {Gottl{\"o}ber}, S. 2006{\natexlab{b}}, \mnras, 369, 335

\bibitem[{{Pen}(1998)}]{pen98}
{Pen}, U.-L. 1998, \apj, 504, 601

\bibitem[{{Pisani} {et~al.}(2015){Pisani}, {Sutter}, {Hamaus}, {Alizadeh},
  {Biswas}, {Wandelt}, \& {Hirata}}]{pisani+15}
{Pisani}, A., {Sutter}, P.~M., {Hamaus}, N., {et~al.} 2015, \prd, 92, 083531

\bibitem[{{Platen} {et~al.}(2007){Platen}, {van de Weygaert}, \&
  {Jones}}]{platen&jones07}
{Platen}, E., {van de Weygaert}, R., \& {Jones}, B. J.~T. 2007, \mnras, 380,
  551

\bibitem[{{Pollina} {et~al.}(2017){Pollina}, {Hamaus}, {Dolag}, {Weller},
  {Baldi}, \& {Moscardini}}]{pollina+17}
{Pollina}, G., {Hamaus}, N., {Dolag}, K., {et~al.} 2017, \mnras, 469, 787

\bibitem[{{Pontzen} {et~al.}(2016){Pontzen}, {Slosar}, {Roth}, \&
  {Peiris}}]{pontzen+16}
{Pontzen}, A., {Slosar}, A., {Roth}, N., \& {Peiris}, H.~V. 2016, \prd, 93,
  103519

\bibitem[{{Scoccimarro}(1998)}]{scoccimarro98}
{Scoccimarro}, R. 1998, \mnras, 299, 1097

\bibitem[{{Seljak} {et~al.}(2009){Seljak}, {Hamaus}, \&
  {Desjacques}}]{seljak+09}
{Seljak}, U., {Hamaus}, N., \& {Desjacques}, V. 2009, \prl, 103, 091303

\bibitem[{{Shim} {et~al.}(2015){Shim}, {Lee}, \& {Hoyle}}]{shim+15}
{Shim}, J., {Lee}, J., \& {Hoyle}, F. 2015, \apj, 815, 107

\bibitem[{{Shim} {et~al.}(2021){Shim}, {Park}, {Kim}, \& {Hwang}}]{shim+21}
{Shim}, J., {Park}, C., {Kim}, J., \& {Hwang}, H.~S. 2021, \apj, 908, 211

\bibitem[{{Stopyra} {et~al.}(2021){Stopyra}, {Peiris}, \&
  {Pontzen}}]{stopyra+21}
{Stopyra}, S., {Peiris}, H.~V., \& {Pontzen}, A. 2021, \mnras, 500, 4173

\bibitem[{{Suto} {et~al.}(1984){Suto}, {Sato}, \& {Sato}}]{suto+84}
{Suto}, Y., {Sato}, K., \& {Sato}, H. 1984, Progress of Theoretical Physics,
  71, 938

\bibitem[{{Sutter} {et~al.}(2015){Sutter}, {Lavaux}, {Hamaus}, {Pisani},
  {Wandelt}, {Warren}, {Villaescusa-Navarro}, {Zivick}, {Mao}, \&
  {Thompson}}]{sutter+15}
{Sutter}, P.~M., {Lavaux}, G., {Hamaus}, N., {et~al.} 2015, Astronomy and
  Computing, 9, 1

\bibitem[{{Tegmark} {et~al.}(2004){Tegmark}, {Blanton}, {Strauss}, {Hoyle},
  {Schlegel}, {Scoccimarro}, {Vogeley}, {Weinberg}, {Zehavi}, {Berlind},
  {Budavari}, {Connolly}, {Eisenstein}, {Finkbeiner}, {Frieman}, {Gunn},
  {Hamilton}, {Hui}, {Jain}, {Johnston}, {Kent}, {Lin}, {Nakajima}, {Nichol},
  {Ostriker}, {Pope}, {Scranton}, {Seljak}, {Sheth}, {Stebbins}, {Szalay},
  {Szapudi}, {Verde}, {Xu}, {Annis}, {Bahcall}, {Brinkmann}, {Burles},
  {Castander}, {Csabai}, {Loveday}, {Doi}, {Fukugita}, {Gott}, {Hennessy},
  {Hogg}, {Ivezi{\'c}}, {Knapp}, {Lamb}, {Lee}, {Lupton}, {McKay}, {Kunszt},
  {Munn}, {O'Connell}, {Peoples}, {Pier}, {Richmond}, {Rockosi}, {Schneider},
  {Stoughton}, {Tucker}, {Vanden Berk}, {Yanny}, {York}, \& {SDSS
  Collaboration}}]{tegmark+04}
{Tegmark}, M., {Blanton}, M.~R., {Strauss}, M.~A., {et~al.} 2004, \apj, 606,
  702

\bibitem[{{Tonegawa} {et~al.}(2020){Tonegawa}, {Park}, {Zheng}, {Park}, {Hong},
  {Hwang}, \& {Kim}}]{tonegawa+20}
{Tonegawa}, M., {Park}, C., {Zheng}, Y., {et~al.} 2020, \apj, 897, 17

\bibitem[{{Tully}(2015)}]{tully15}
{Tully}, R.~B. 2015, \aj, 149, 171

\bibitem[{{Vogeley} {et~al.}(1994){Vogeley}, {Park}, {Geller}, {Huchra}, \&
  {Gott}}]{vogeley+94}
{Vogeley}, M.~S., {Park}, C., {Geller}, M.~J., {Huchra}, J.~P., \& {Gott},
  J.~Richard, I. 1994, \apj, 420, 525

\bibitem[{{Wang} {et~al.}(2009){Wang}, {Mo}, {Jing}, {Guo}, {van den Bosch}, \&
  {Yang}}]{wang+09}
{Wang}, H., {Mo}, H.~J., {Jing}, Y.~P., {et~al.} 2009, \mnras, 394, 398

\bibitem[{{Woodfinden} {et~al.}(2022){Woodfinden}, {Nadathur}, {Percival},
  {Radinovic}, {Massara}, \& {Winther}}]{woodfinden+22}
{Woodfinden}, A., {Nadathur}, S., {Percival}, W.~J., {et~al.} 2022, \mnras,
  arXiv:2205.06258

\bibitem[{{Zehavi} {et~al.}(2011){Zehavi}, {Zheng}, {Weinberg}, {Blanton},
  {Bahcall}, {Berlind}, {Brinkmann}, {Frieman}, {Gunn}, {Lupton}, {Nichol},
  {Percival}, {Schneider}, {Skibba}, {Strauss}, {Tegmark}, \&
  {York}}]{Zehavi+11}
{Zehavi}, I., {Zheng}, Z., {Weinberg}, D.~H., {et~al.} 2011, \apj, 736, 59

\bibitem[{{Zel'Dovich}(1970)}]{zeldovich70}
{Zel'Dovich}, Y.~B. 1970, \aap, 500, 13

\end{thebibliography}
\end{CJK}
\end{document}